\documentclass[conference]{IEEEtran}
\IEEEoverridecommandlockouts

\usepackage{cite}
\usepackage{amsmath,amssymb,amsfonts}
\usepackage{algorithmic}
\usepackage{graphicx}
\usepackage{textcomp}
\usepackage{xcolor}
\usepackage{booktabs}
\usepackage{multirow}
\usepackage{array}  

\usepackage{wasysym}

\usepackage[top=0.75in, bottom=1.09in, left=0.65in, right=0.65in]{geometry}

\def\BibTeX{{\rm B\kern-.05em{\sc i\kern-.025em b}\kern-.08em
    T\kern-.1667em\lower.7ex\hbox{E}\kern-.125emX}}

\begin{document}


\title{MV-Gate: Insider Threat Detection via Multi-View Behavioral Statistics and Semantic Modeling}


\author{
\IEEEauthorblockN{
Kaichuan Kong\textsuperscript{a}, 
Dongjie Liu\textsuperscript{a,}\IEEEauthorrefmark{1}, 
Xiaobo Jin\textsuperscript{b}, 
Guanggang Geng\textsuperscript{a}}
\thanks{\IEEEauthorrefmark{1} Corresponding author. }
\IEEEauthorblockA{
\textsuperscript{a}College of Cyber Security, Jinan University, Guangzhou, China\\
\textsuperscript{b}School of Advanced Technology, Xi’an Jiaotong-Liverpool University, Suzhou, China\\
\ willkkc@stu2021.jnu.edu.cn, \{djliu, gggeng\}@jnu.edu.cn, xiaobo.jin@xjtlu.edu.cn \\
}
}

\maketitle

\begin{abstract}
Insider threats often reveal early anomalies through disruptions in behavioral statistics—such as altered recurrence patterns or short- versus long-term frequency shifts—rather than changes in event semantics. Yet, as the field has shifted from statistical modeling to log tokenization and deep sequential encoders, these statistical cues are weakened or lost, leaving current models insensitive to gradual and low-visibility insider behaviors.
We propose MV-Gate, a multi-view behavior modeling framework that explicitly integrates statistical regularities with sequence semantics. MV-Gate constructs three aligned behavioral sequences: activity tokens, multi-scale status signals capturing recurrence patterns, and frequency-deviation signals describing short- vs long-term intensity differences. An anomaly-aware gating mechanism injects these statistical views into the attention computation, guiding the encoder to emphasize statistically irregular events.
Experiments on CERT r4.2, CERT r5.2, and ADFA-LD show that MV-Gate achieves notable gains over classical, deep-learning, and domain-specific baselines, particularly for progressive, weak-signal threats. These results highlight the necessity of jointly modeling statistical and sequential evidence for robust insider-threat detection.
\end{abstract}


\section{Introduction}
Insider threats remain a major security challenge because malicious actions are performed by legitimate users whose behavior closely resembles normal operations. Such attacks progress slowly and generate weak, temporally sparse signals~\cite{alzaabi2024review}, leading to long detection delays. Industry reports including the 2025 Ponemon Institute study~\cite{ponemon2025} show that many insider incidents remain unnoticed for weeks or months, highlighting the need to detect subtle behavioral deviations rather than explicit malicious activities.

A key observation is that early insider anomalies rarely stem from the semantics of an individual event, but instead from disruptions in a user’s \emph{behavioral statistics}—whether an action is new, rare, resurfacing after inactivity, or showing abnormal short-long-term frequency shifts. Earlier insider-threat detection (ITD) systems relied on statistical features and temporal window analysis, making them naturally sensitive to such irregularities.

With the field’s shift toward deep log-sequence modeling, a methodological gap has emerged. Modern approaches tokenize logs and process them with Long Short-Term Memory (LSTM) networks, Gated Recurrent Units (GRUs), or Transformer architectures~\cite{vaswani2017attention, he2021insider, huang2021itdbert}. While these models capture semantic and contextual patterns, their \emph{token-only} representations discard statistical regularities such as recurrence structure or frequency deviation. Consequently, they model \emph{what} happened but remain insensitive to \emph{how abnormally often} behaviors occur—signals that frequently precede weak-signal or gradually evolving insider threats.

To address this gap, we propose \textbf{MV-Gate} (Multi-View Gated Attention), a lightweight framework that integrates statistical behavioral cues with sequential modeling. MV-Gate constructs three aligned sequences: (i) activity tokens encoding event semantics, (ii) a \emph{status view} capturing multi-scale recurrence patterns, and (iii) a \emph{frequency-deviation view} quantifying short- versus long-term intensity changes. A \textbf{dynamic anomaly-aware gating mechanism} injects these statistical cues into the attention computation, enabling the model to emphasize irregular events without modifying the underlying Transformer structure.

Our contributions are:
\begin{itemize}

    \item \textbf{Multi-View Activity Representation:} A lightweight statistical feature based on three views (semantic tokens, status, and frequency) and semantic embedding is jointly optimized to avoid feature engineering dependencies.
    

    \item \textbf{Anomaly-Aware Gating Mechanism:} A dynamic risk-aware gating mechanism is designed to adjust attention weights based on statistical anomaly scores, enabling the model to focus on highly suspicious events and overcome the insensitivity of traditional attention to statistical signals.

    \item \textbf{End-to-End Detection Framework:} A gated Transformer-based architecture integrating multi-view fusion and anomaly scoring in an end-to-end pipeline.
    
    \item \textbf{Extensive Evaluation:} Experiments on CERT r4.2, CERT r5.2, and ADFA-LD show significant improvements over classical, deep learning, and domain-specific baselines.
\end{itemize}

The remainder of the paper reviews related work (Section~\ref{sec:related}), formalizes the problem (Section~\ref{sec:problem}), presents the MV-Gate model (Section~\ref{sec:method}), and reports experiments (Section~\ref{sec:exp}).

\vspace{-0.3em}
\section{Related Work}
\label{sec:related}

Insider threat detection (ITD) spans statistical-learning methods and modern deep sequence models. Despite substantial progress, the shift toward token-only representations has weakened the modeling of behavioral irregularities such as recurrence and frequency deviation. We review related work and summarize this gap below.

\subsection{Classical Machine-Learning Approaches}

Early ITD systems primarily relied on handcrafted statistical features derived from enterprise audit logs, including login frequencies, device usage patterns, and file-access statistics. Liu~\emph{et al.}~\cite{liu2019log2vec} proposed Log2vec, which embeds heterogeneous user--file--device interactions using graph-based representations. Le~\emph{et al.}~\cite{le2020analyzing} evaluated classical supervised classifiers such as Logistic Regression (LR), Random Forest (RF), and Extreme Gradient Boosting (XGBoost), demonstrating that carefully designed statistical features can form strong baselines. Feature selection methods, including Information Gain (IG) and Correlation-Based Feature Selection (CFS)~\cite{bin2022insider}, further improve generalization by removing redundant dimensions.
Unsupervised anomaly detection methods have also been widely studied. Bartoszewski~\emph{et al.}~\cite{bartoszewski2021anomaly} compared Local Outlier Factor (LOF), One-Class Support Vector Machine (OC-SVM), Isolation Forest (IF), and Hidden Markov Models (HMMs). Le~\emph{et al.}~\cite{le2021anomaly} applied autoencoder reconstruction to high-dimensional behavioral vectors, while Yousef~\emph{et al.}~\cite{yousef2023machine} used IF for session-level anomaly scoring. These classical approaches retain explicit behavioral statistics and are therefore sensitive to changes in frequency, recurrence, or intensity. However, they lack the ability to model long-range temporal dependencies or semantic context, limiting their performance on complex behavior sequences.

\subsection{Deep Learning-Based Behavioral Modeling}

Deep learning approaches improve ITD by directly learning sequential patterns from raw event streams. Recurrent neural networks such as Long Short-Term Memory (LSTM) and Gated Recurrent Unit (GRU) models~\cite{he2021insider,pal2023temporal} capture short-term transitions, while Transformer-based architectures~\cite{huang2021itdbert,xiao2024unveiling} leverage self-attention to model long-range dependencies without manual feature engineering.
Beyond sequential encoders, several alternative representations have been explored. Budžys~\emph{et al.}~\cite{budvzys2024deep} transformed keystroke dynamics into time--frequency images for Convolutional Neural Network (CNN) analysis, while Gayathri~\emph{et al.}~\cite{gayathri2024spcagan} introduced the Synthetic Principal Component Analysis Generative Adversarial Network (SPCAGAN) to alleviate dataset imbalance through synthetic sample generation. Structural modeling has also gained traction: Xiao~\emph{et al.}~\cite{xiao2022robust} and Roy~\emph{et al.}~\cite{roy2024graphch} applied Graph Neural Networks (GNNs) to capture user--resource interaction structures, and Cai~\emph{et al.}~\cite{cai2024lan} integrated graph embeddings with chronological behavior modeling.
More recently, Large Language Models (LLMs) have emerged as an alternative paradigm for log analysis. LogGPT~\cite{qi2023loggpt} and LogPrompt~\cite{liu2024interpretable} employ prompt-based inference for zero-shot or few-shot anomaly detection. While capable of semantic reasoning, LLM-based approaches incur high computational cost and, like other deep models, operate on tokenized event sequences that lack explicit statistical representations of user behaviors.



\subsection{Research Gap}
Despite recent progress, most ITD systems rely on \emph{tokenized activity sequences}, treating all events uniformly and overlooking two early statistical indicators of insider incidents: (i) multi-scale recurrence patterns (e.g., new, rare, or resurfacing actions) and (ii) short--long term frequency deviations (e.g., bursts or suppressed behaviors). Although LSTM, GRU, and Transformer encoders may capture these signals implicitly, they offer neither explicit statistical modeling nor interpretable cues for abnormality. Other research directions, such as GNN-based structural modeling, generative augmentation, and time--frequency representations, address orthogonal challenges and similarly lack mechanisms for integrating statistical priors with temporal reasoning. 

To date, no ITD framework jointly represents multi-scale behavioral status and frequency deviation or injects such statistical cues into attention computation. This limitation motivates MV-Gate, which couples explicit multi-view behavioral statistics with anomaly-aware gated attention to emphasize actions that deviate from normal behavioral rhythms.

\begin{figure*}[!ht]
  \centering
  \includegraphics[width=0.93\textwidth]{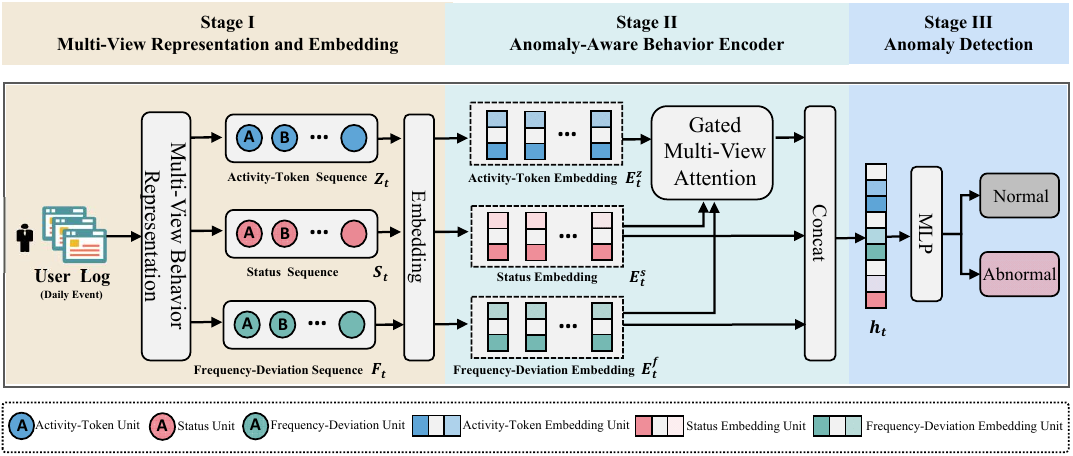}
    \caption{
    MV-Gate framework. Logs are converted into three aligned views 
    (token, status, and frequency). Status and frequency cues produce anomaly-aware 
    gates that modulate token-level attention. The gated representation is fused 
    and used for final anomaly detection.
    }
  \label{fig:Structure}
\vspace{-0.5cm}  
\end{figure*}

\vspace{-0.5em}
\section{Problem Formulation}
\label{sec:problem}
\vspace{-0.2em}

Enterprise audit logs form a chronological sequence of user events,
where each event is represented as
\begin{equation}
x_t = (u_t, b_t, \tau_t),
\end{equation}
with $u_t$ denoting the user, $b_t$ the behavior category, and $\tau_t$
the timestamp. For an analysis window, the event sequence is
$X=\{x_1,\dots,x_T\}$.

Following standard practice~\cite{huang2021itdbert}, each event is mapped
to a discrete activity token through
\begin{equation}
z_t=f_{\mathrm{map}}(b_t,\tau_t), \qquad z_t\in\mathcal{V},
\end{equation}
yielding the activity-token sequence $Z=\{z_t\}$.

From $Z$, we further derive two auxiliary statistical behavioral views:
(i) a multi-scale status view $S=\{s_t\}$ encoding recurrence indicators,
and (ii) a frequency-deviation view $F=\{f_t\}$ capturing short--long
term intensity shifts. Both signals are computed causally (using only
past observations) and temporally aligned with $Z$.

Given the synchronized triplet $(Z,S,F)$, insider threat detection is
formulated as sequence-level anomaly scoring:
\begin{equation}
\hat{y}=f_\theta(Z,S,F), \qquad \hat{y}\in[0,1],
\end{equation}
where $\hat{y}$ denotes the predicted maliciousness and $y\in\{0,1\}$
are binary supervision labels.

\section{Proposed Method}
\label{sec:method}

This section presents MV-Gate, a lightweight behavior modeling framework
that enhances sequence encoders with multi-view behavioral cues and
anomaly-aware attention modulation. The model consists of three main
components: (1) multi-view behavior representation, (2) anomaly-aware
attention gating, and (3) a gated Transformer encoder for sequence-level
classification. Figure~\ref{fig:Structure} provides an overview.

\begin{figure}[!t] 
\centering 
\includegraphics[width=0.95\linewidth]{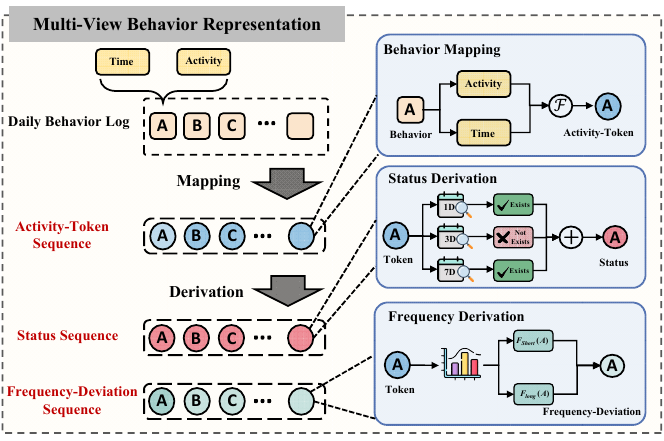} 
\caption{
Overview of the multi-view behavior representation. Each log event is first
tokenized, then complemented with a multi-scale status view and a short--long
term frequency-deviation view, forming three synchronized behavior sequences.
}
\label{fig:Stage1} 
\vspace{-0.5cm}  
\end{figure}

\subsection{Multi-View Representation and Embedding}
\label{sec:multiview}

MV-Gate operates on three synchronized behavioral views derived from the
tokenized activity sequence: (i) an activity-token view $Z$, (ii) a
multi-scale status view $S$, and (iii) a frequency-deviation view $F$.
All views share the same temporal index and are later embedded jointly
through a unified embedding layer.

\subsubsection{Multi-View Behavior Generation}

\paragraph{Activity Token View}
Each raw log event $x_t$ is converted into a discrete activity token 
that encodes its semantic behavior type and coarse temporal context, 
following the standard tokenization procedure widely adopted in 
log-based anomaly detection~\cite{huang2021itdbert}. Formally,
\begin{equation}
    z_t = f_{\mathrm{map}}(b_t, \tau_t), \qquad z_t \in \mathcal{V},
\end{equation}
which produces the activity-token sequence
\begin{equation}
    Z = \{z_t\}_{t=1}^{T}.
\end{equation}

This symbolic abstraction provides a unified representation of heterogeneous log events while preserving their chronological structure.

\paragraph{Status View}
Recurrence irregularity is a key indicator of insider threats, as newly
emerging or rarely occurring behaviors often carry higher risk than
routine actions. To quantify how unusual the current token $z_t$ is, we
check whether it has appeared within $K$ historical windows of sizes
$\{w_1,\dots,w_K\}$. For the $k$-th window,
\begin{equation}
s_t^{(k)} = \mathbf{I}\!\left(z_t \notin Z_{(t-w_k):(t-1)}\right),
\end{equation}
where $\mathbf{I}(\cdot)$ is an indicator function and  $s_t^{(k)}=1$ indicates that $z_t$ has not appeared in the past $w_k$ steps. The multi-scale rarity level is then computed as
\begin{equation}
s_t = \sum_{k=1}^{K} s_t^{(k)}.
\end{equation}

A larger $s_t$ reflects a rarer or newly emerging behavior. For example,
under window sizes $\{1,3,7\}$, if $z_t$ was absent in the last $3$ and $7$
steps but present in the last $1$, then $s_t^{(k)}=[0,1,1]$ and $s_t=2$.

\paragraph{Frequency View}
To characterize short--long term behavioral shifts, we compare the
normalized short-term and long-term occurrence frequencies of token
$z_t$ under window lengths $h_s$ and $h_l$ ($h_s \ll h_l$):
\begin{equation}
\begin{aligned}
f_t^{(s)} &= \frac{\mathrm{count}(z_t \mid Z_{(t-h_s):(t-1)})}{h_s},\\[4pt]
f_t^{(l)} &= \frac{\mathrm{count}(z_t \mid Z_{(t-h_l):(t-1)})}{h_l}.
\end{aligned}
\end{equation}

The frequency-deviation view is defined as their relative contrast:
\begin{equation}
f_t = \frac{f_t^{(s)} - f_t^{(l)}}{f_t^{(l)} + \epsilon}.
\end{equation}
where $\epsilon$ is a small constant added for numerical stability to avoid
division by zero when $f_t^{(l)}$ is extremely small.
For example, if a token appears $3$ times in the last $3$ days and
$10$ times in the last $30$ days, then $f_t^{(s)}=1.0$, $f_t^{(l)}\approx0.33$,
and $f_t\approx2$, indicating a recent surge over the long-term baseline.

\subsubsection{Sequence Embedding}
\label{sec:embedding}

The three behavioral views $(Z,S,F)$ share the same temporal length but
have heterogeneous formats. Each view is projected into a shared
$d$-dimensional space by a view-specific embedding function:
\begin{equation}
e_t^{z}=E_z(z_t),\qquad
e_t^{s}=E_s(s_t),\qquad
e_t^{f}=E_f(f_t).
\end{equation}
Here, $E_z$ is an embedding lookup for activity tokens and $E_s,E_f$ are
linear projections for the scalar status and frequency-deviation values.
The corresponding embedding sequences are $E^{z},E^{s},E^{f}.$

\subsection{Anomaly-Aware Behavior Encoder}
\label{sec:encoder}

The activity-token view serves as the primary behavioral stream and is
encoded by a Transformer encoder. To incorporate auxiliary statistical
signals, we introduce an anomaly-aware gating mechanism that modulates
the attention computation based on the status and frequency embeddings.

\paragraph{Risk Gate}
For each timestep, a scalar gate is computed to indicate the
abnormality level of the event:
\begin{equation}
g_t = \sigma\!\left(W_g \left[\, e_t^s \Vert e_t^f \,\right] + b_g \right),
\qquad g_t \in (0,1).
\end{equation}

\paragraph{Gated Self-Attention}
Given token embeddings $e_t^{z}$, the standard attention projections are:
\begin{equation}
Q_t = W_Q e_t^{z}, \quad
K_t = W_K e_t^{z}, \quad
V_t = W_V e_t^{z}.
\end{equation}
Anomaly-awareness is injected by scaling the key vectors:
\begin{equation}
K'_t = g_t \cdot K_t.
\end{equation}
The gated attention logits are computed as:
\begin{equation}
L_{i,t} = \frac{Q_i \left(K'_t\right)^{\top}}{\sqrt{d}}.
\end{equation}
Except for this modulation, the encoder follows the standard Transformer
design (multi-head attention, feed-forward layers, residual connections,
and layer normalization). The output is a sequence of contextualized
token representations:
\begin{equation}
\left\{ h_t^{\mathrm{tok}} \right\}_{t=1}^{T}.
\end{equation}

\subsection{Anomaly Detection}
\label{sec:detection}

A learnable classification token $[\mathrm{CLS}]$ is prepended to the
activity-token sequence, and its final hidden state provides a
sequence-level summary:
\begin{equation}
h_{\mathrm{tok}} = h_{\mathrm{CLS}}.
\end{equation}

The auxiliary views are aggregated via mean pooling:
\begin{equation}
\begin{aligned}
h_{\mathrm{sta}}  &= \mathrm{MeanPool}\!\left( \{ e_t^s \} \right),
h_{\mathrm{freq}} &= \mathrm{MeanPool}\!\left( \{ e_t^f \} \right).
\end{aligned}
\end{equation}

The fused representation is:
\begin{equation}
h = \left[\, h_{\mathrm{tok}} \;\Vert\; h_{\mathrm{sta}} \;\Vert\; h_{\mathrm{freq}} \,\right].
\end{equation}

An MLP classifier produces the anomaly probability:
\begin{equation}
\hat{y} = \sigma\!\left( \mathrm{MLP}(h) \right).
\end{equation}
The model is trained using binary cross-entropy:
\begin{equation}
\mathcal{L}
= - y \log(\hat{y}) - \left(1-y\right)\log(1-\hat{y}),
\end{equation}
where $y$ is the ground truth.

\begin{table*}[!t]
    \footnotesize
    \renewcommand{\arraystretch}{0.9}
    \centering
    \renewcommand{\arraystretch}{0.8}   
    \caption{Performance comparison across CERT r4.2, CERT r5.2, and ADFA-LD. 
    Best results in \textbf{bold}, second-best \underline{underlined}.}
    \label{table:three-dataset-combined}
    \begin{tabular}{llcccc|cccc|cccc}
    \toprule
    \multirow{2}{*}{\bfseries Category} &
    \multirow{2}{*}{\bfseries Method} &
        \multicolumn{4}{c}{\bfseries CERT r4.2} &
        \multicolumn{4}{c}{\bfseries CERT r5.2} &
        \multicolumn{4}{c}{\bfseries ADFA-LD} \\
    \cmidrule(lr){3-6} \cmidrule(lr){7-10} \cmidrule(lr){11-14}
    & & Rec$\uparrow$ & Prec$\uparrow$ & Acc$\uparrow$ & F1$\uparrow$
      & Rec$\uparrow$ & Prec$\uparrow$ & Acc$\uparrow$ & F1$\uparrow$
      & Rec$\uparrow$ & Prec$\uparrow$ & Acc$\uparrow$ & F1$\uparrow$  \\
    \midrule

    \multirow{5}{*}{Classical }
    & LR           
        & 0.688 & 0.958 & 0.953 & 0.783  
        & 0.649 & 0.935 & 0.944 & 0.748
        & 0.891 & 0.947 & 0.940 & 0.918 \\

    & RF           
        & 0.769 & \textbf{0.986} & 0.968 & 0.838  
        & 0.721 & 0.951 & 0.961 & 0.806
        & 0.902 & \underline{0.954} & \underline{0.951} & \underline{0.927} \\
        
    & XGBoost      
    & 0.827 & 0.957 & 0.973 & 0.871  
    & 0.854 & \underline{0.973} & \underline{0.978} & 0.899
    & \underline{0.931} & 0.922 & 
      0.943 & 0.926 \\

    & CNN          
    & 0.410 & 0.867 & 0.923 & 0.539  
    & 0.584 & 0.924 & 0.940 & 0.692
    & 0.594 & 0.882 & 0.815 & 0.710 \\

    & Transformer  
        & 0.779 & 0.944 & 0.962 & 0.839  
        & 0.781 & 0.876 & 0.953 & 0.805
        & 0.743 & 0.824 & 0.842 & 0.781 \\

    \midrule

    \multirow{4}{*}{Domain-Specific }

    & ITDBERT      
        & 0.884 & 0.912 & 0.960 & 0.898  
        & 0.889 & 0.914 & 0.961 & 0.901
        & 0.713 & 0.878 & 0.853 & 0.787 \\

    & CATE         
        & 0.904 & 0.936 & \underline{0.980} & \underline{0.911}  
        & 0.893 & 0.972 & 0.983 & \underline{0.926}
        & 0.842 & 0.817 & 0.868 & 0.829 \\

    & ITDLM        
        & 0.852 & 0.906 & 0.950 & 0.879  
        & \underline{0.930} & 0.843 & 0.951 & 0.884
        & 0.727 & 0.920 & 0.861 & 0.812 \\

    & LogGPT       
        & \underline{0.920} & 0.880 & 0.959 & 0.899  
        & 0.925 & 0.891 & 0.963 & 0.907
        & 0.696 & 0.889 & 0.865 & 0.780 \\
    \midrule

    \multirow{1}{*}{\bfseries Ours}
    & \textbf{MV-Gate }
        & \textbf{0.927} & \underline{0.985} & \textbf{0.988} & \textbf{0.951}
        & \textbf{0.934} & \textbf{0.991} & \textbf{0.991} & \textbf{0.957}
        & \textbf{0.960} & \textbf{0.970} & \textbf{0.974} & \textbf{0.965} \\
    \bottomrule
\end{tabular}
\vspace{-0.5cm}  
\end{table*}

\begin{table}[!t]
    \footnotesize
    \renewcommand{\arraystretch}{0.8}
    \centering
    \caption{Ablation study on CERT r4.2. 
    Each variant removes one component from the full model. 
    Best results in \textbf{bold}.}
    \label{tab:ablation}
    \begin{tabular}{lcccc}
    \toprule
    \bfseries Variant & Rec$\uparrow$ & Prec$\uparrow$ & Acc$\uparrow$ & F1$\uparrow$ \\

    \midrule
    w/o Status View 
        & 0.876 & 0.973 & 0.974 & 0.918 \\
    w/o Frequency View 
        & 0.886 & 0.995 & 0.984 & 0.931 \\
    w/o Multi-View Fusion\,(Freq+Status) 
        & 0.805 & 0.973 & 0.971 & 0.866 \\
    w/o Gated Attention 
        & 0.913 & 0.989 & 0.983 & 0.934 \\
        \midrule
    \textbf{Full Model (Ours)} 
        & \textbf{0.927} & \textbf{0.985} & \textbf{0.988} & \textbf{0.951} \\
    \bottomrule
    \end{tabular}
\vspace{-0.5cm}  
\end{table}

\vspace{-0.5em}
\section{Experimental Configuration}
\label{sec:exp}
\vspace{-0.2cm}  

\vspace{-0.5em}
\subsection{Datasets}
\vspace{-0.2cm}

We evaluate the proposed MV-Gate model on three public datasets:
the \emph{Carnegie Mellon University CERT Insider Threat Dataset}
(versions r4.2 and r5.2)~\cite{lindauer2020insider},
and the \emph{Australian Defence Force Academy Linux Dataset} (ADFA-LD)~\cite{creech2013semantic}.

For CERT datasets, we follow Scenario~2 and use the officially provided users with labeled insider-threat activities. Daily enterprise logs (e.g., logon, email, web, file, and device events) are grouped into behavior sessions according to temporal continuity. 
For ADFA-LD, all intrusive traces are merged into a single anomaly class, and raw system-call sequences are transformed using the same feature pipeline as CERT.

\subsection{Implementation Settings}
All datasets are processed using the unified multi-view pipeline in Section~\ref{sec:multiview}.
Each session is converted into an activity-token sequence $Z$, and its status view $S$ and frequency-deviation view $F$ are derived using past events only.
The synchronized views $(Z,S,F)$ are embedded jointly and fed into MV-Gate.
We use an 80/20 train–test split and apply 5-fold cross-validation on the training set for hyperparameter tuning.
All baselines receive the same tokenized activity sequences for fair comparison, and their hyperparameters follow recommended configurations in prior work.
For ADFA-LD, system-call traces are treated as activity tokens and processed using the same multi-view derivation procedure.

\subsection{Baseline Methods}
We compare MV-Gate against representative baselines widely used in log-based anomaly detection. 
\textbf{(1) Classical models.} These include general-purpose detectors based on statistical features or conventional deep models, such as Logistic Regression (LR), Random Forest (RF), XGBoost (XGB), CNN classifiers, and Transformer-based sequence encoders~\cite{bartoszewski2021anomaly, le2021anomaly}. 
\textbf{(2) Domain-specific models.} These methods are tailored for insider-threat or log-sequence analysis and incorporate temporal or semantic behavior modeling. We evaluate ITDBERT~\cite{huang2021itdbert}, CATE~\cite{xiao2024unveiling}, ITDLM~\cite{song2025confront}, and LogGPT~\cite{qi2023loggpt}, covering representative supervised and LLM-based reasoning approaches.

\subsection{Evaluation Metrics}
Detection performance is assessed using four standard metrics: Recall, Precision, Accuracy, and F1-score. 
Recall measures the system’s ability to detect true threats, Precision evaluates the correctness of predicted anomalies, and F1-score provides a balanced assessment under the class imbalance typically seen in insider-threat datasets.

\section{Results and Discussions}
\label{sec:results}

\subsection{Baseline Comparison}






As shown in Table \ref{table:three-dataset-combined}, MV-Gate consistently achieves the best overall performance across all three datasets. Compared with classical baselines (LR, RF, XGBoost) and generic deep models (CNN, Transformer), our method yields substantial gains in both recall and F1, indicating that multi-view behavioral cues are more discriminative than single-view sequence features. Particularly on enterprise-scale CERT r4.2/5.2, MV-Gate maintains high accuracy while significantly reducing missed anomalies, demonstrating superior robustness under imbalanced real-world log distributions.

Moreover, domain-specific detectors such as ITDBERT, CATE, ITDLM, and LogGPT also fall short of MV-Gate. As shown in Table \ref{table:three-dataset-combined}, our model achieves the highest performance on every key metric, confirming the value of anomaly-aware gating for integrating heterogeneous behavioral views. These results verify that MV-Gate provides a stronger and more reliable representation of insider-threat patterns than existing handcrafted or sequence-only architectures.

\subsection{Ablation Study}
To systematically assess the contribution of each model component, we conduct the ablation study exclusively on CERT r4.2, as reported in Table~\ref{tab:ablation}. CERT r4.2 provides a stable user population, rich multi-view behavioral logs, and well-structured insider threat scenarios, making it the most suitable benchmark for isolating the effects of individual architectural modules. As shown in the table, removing any component leads to a notable performance degradation, with the largest drop observed when disabling multi-view fusion or gated attention. These results confirm that both the multi-view behavior representations and the anomaly-aware gating mechanism play essential roles in boosting detection accuracy, thereby validating the design principles of our proposed model.

\subsection{Hyperparameter Study}

\begin{figure}[t]
  \centering

  \begin{minipage}[b]{0.45\linewidth}
    \centering
    \includegraphics[width=0.95\linewidth]{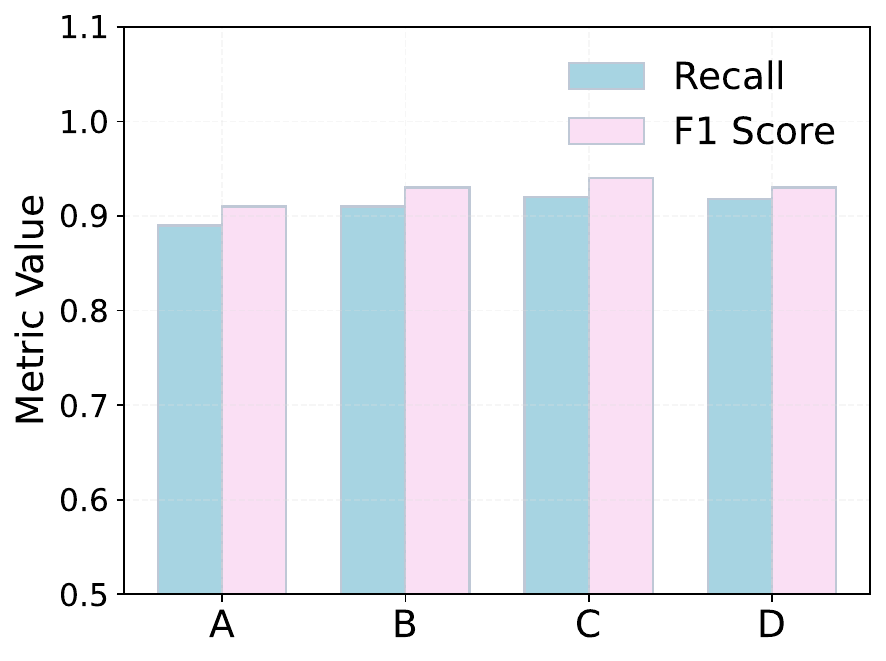}
    {\footnotesize (a) Status Window Size}
  \end{minipage}
  \begin{minipage}[b]{0.45\linewidth}
    \centering
    \includegraphics[width=0.95\linewidth]{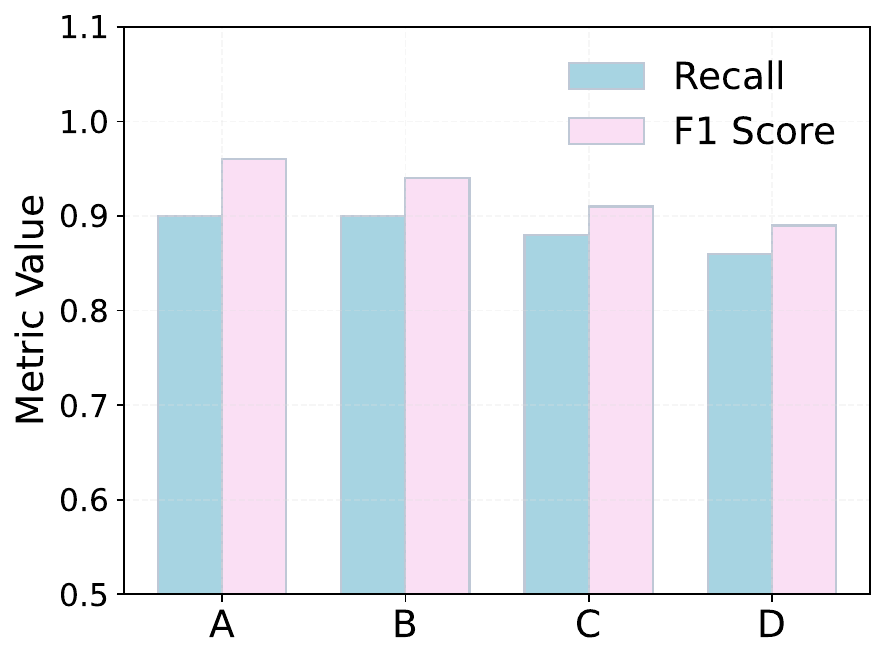}
    {\footnotesize (b) Frequency Window Pair}
  \end{minipage}

  \caption{\footnotesize Effect of statistical-view configurations on Recall and F1.}
  \label{fig:stat_hyperparams}
\vspace{-0.3cm}  
\end{figure}

\begin{figure}[t]
  \centering

  \begin{minipage}[b]{0.45\linewidth}
    \centering
    \includegraphics[width=0.95\linewidth]{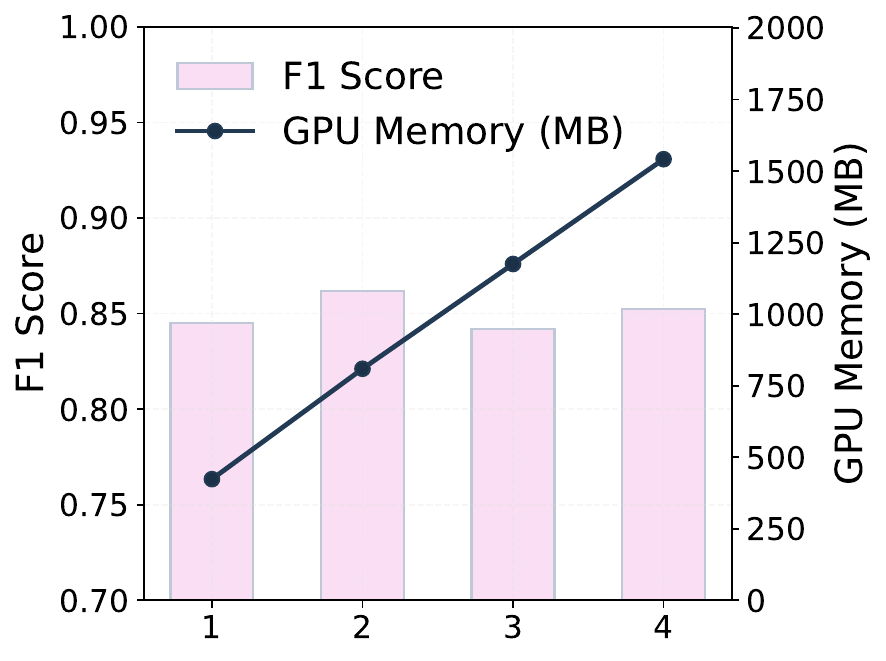}
    {\footnotesize (a) Attention Encoder Layers}
  \end{minipage}
  \begin{minipage}[b]{0.45\linewidth}
    \centering
    \includegraphics[width=0.95\linewidth]{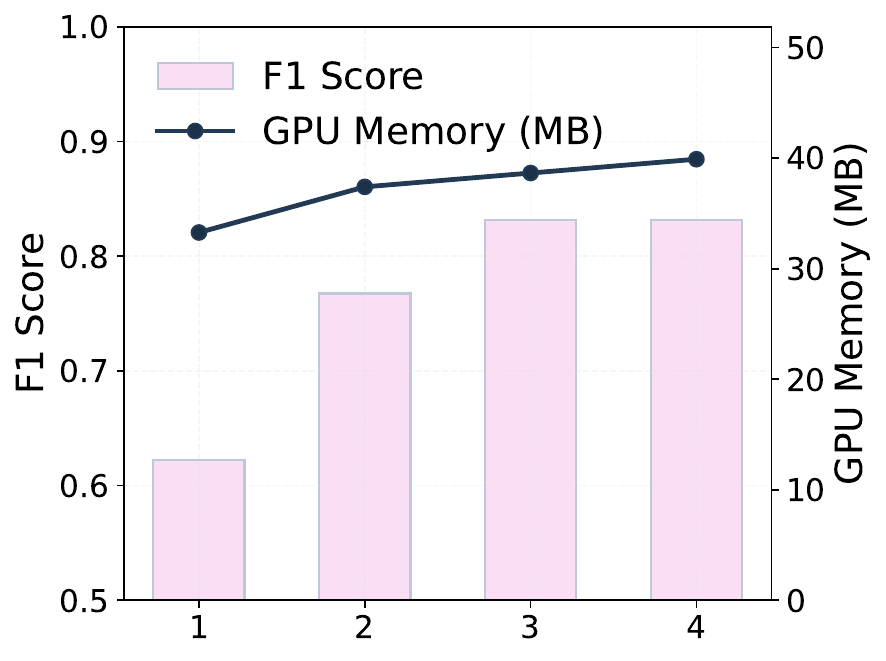}
    {\footnotesize (b) MLP Classifier Layers}
  \end{minipage}

  \caption{\footnotesize Effect of architectural depth on performance and memory usage.}
  \label{fig:arch_sensitivity}
\vspace{-0.5cm}  
\end{figure}

\paragraph{Status Window Size}
Four recurrence configurations are examined:
A: $\{1,3,7\}$,\,
B: $\{1,7,15\}$,\,
C: $\{3,7,15\}$,\,
D: $\{7,15,30\}$.
As shown in Fig.~\ref{fig:stat_hyperparams}(a), the overall trend remains
stable across settings, while short--mid spans (A and C) consistently yield
stronger Recall and F1. These settings are more responsive to rapid yet
recurrent behavioral fluctuations, whereas long-only windows (D) dampen
short-term signals and incur a slight performance drop. Among the four,
configuration C provides the most balanced behavior and is therefore used
as the default.

\paragraph{Frequency Window Pair}
The comparison of short--long frequency pairs
A: $(1,7)$,\,
B: $(3,15)$,\,
C: $(7,15)$,\,
D: $(15,30)$
is presented in Fig.~\ref{fig:stat_hyperparams}(b).
Pairs with a stronger contrast between short and long spans (A and B)
produce more distinct deviation cues, resulting in marginally higher F1.
In contrast, excessively long windows (D) oversmooth the fluctuation
patterns, slightly reducing sensitivity to bursty behaviors. Given its
stable performance, pair A is selected as the default design.

\subsection{Architectural Sensitivity}

\paragraph{Transformer Encoder Depth}
Encoder depth is varied from 1 to 4 layers.
Fig.~\ref{fig:arch_sensitivity}(a) indicates that deeper attention stacks
bring only marginal improvements while incurring a substantial increase in
memory consumption (423~MB→1.5~GB). This phenomenon suggests that ITD
sequences, typically short and sparse, do not benefit from deep attention
hierarchies. A compact 2-layer encoder thus offers the most favorable
efficiency–performance balance.

\paragraph{MLP Head Depth}
MLP classifiers with 1--4 layers are also evaluated.
As shown in Fig.~\ref{fig:arch_sensitivity}(b), performance improves
notably when increasing from 1 to 2 layers and then plateaus, while memory
usage remains almost unchanged (33--40~MB). A 3-layer head provides
sufficient expressive capacity without unnecessary overhead and is adopted
as the final configuration.

\section{Conclusion}
\label{sec:conclusion}

This work introduced MV-Gate, a lightweight multi-view framework that restores statistical behavioral cues often lost in modern token-only insider threat detection pipelines. By combining token semantics, multi-scale status, and frequency-deviation features through anomaly-aware gated attention, MV-Gate effectively captures both routine patterns and weak-signal anomalies. Experiments on CERT r4.2, CERT r5.2, and ADFA-LD confirm substantial gains over classical and deep models. Future work includes the potential for fusion of multimodal logs (such as network traffic + file operations) or integration with LLM to generate semantic interpretations and MV-Gate to provide statistical evidence).

\bibliographystyle{ieeetr}

\end{document}